\documentclass[
    reprint,
    showkeys,
    amsmath,amssymb,
    aps,
    prl,
    floatfix,
]{revtex4-2}

\usepackage{preamble}

\DeclareMathOperator{\abs}{abs}

\begin{document}

\title{Optimal cross-correlation technique to search for
\titletoken{\\} strongly lensed gravitational waves}

\author{Anirban Kopty\orcidlink{0009-0002-8933-9547}}%
    \email{anirban.kopty@iucaa.in}

\author{Sanjit Mitra\orcidlink{0000-0002-0800-4626}}%
    \email{sanjit@iucaa.in}%

\author{Anupreeta More\orcidlink{0000-0001-7714-7076}}%
    \email{anupreeta@iucaa.in}

\affiliation{%
 Inter-University Centre for Astronomy and Astrophysics, Pune
}%

\date{\today}

\begin{abstract}
    As the number of detected gravitational wave (GW) events increases with the improved sensitivity of the observatories, detecting strongly lensed pairs of events is becoming a real possibility. Identifying such lensed pairs, however, remains challenging due to the computational cost and/or the reliance on prior knowledge of source parameters in existing methods. This study investigates a novel approach, Optimal Cross-Correlation Analysis for Multiplets (OCCAM), applied to strain data from one or more detectors for Compact Binary Coalescence (CBC) events identified by GW searches, using an optimal, mildly model-dependent, low computation cost approach to identify strongly lensed candidates. This technique efficiently narrows the search space, allowing for more sensitive, but (much) higher latency, algorithms to refine the results further.
    We demonstrate that our method performs significantly better than other computationally inexpensive methods.
    In particular, we achieve $97\% \, (80\%)$ lensed event detection at a pairwise false positive probability of $\sim$$13\% \, (7\%)$ for a single detector with LIGO design sensitivity, assuming an SNR $\geqslant 10$ astrophysically motivated lensed and unlensed populations.
    Thus, this method, using a network of detectors and in conjunction with sky-localisation information, can enormously reduce the false positive probability, making it highly viable to efficiently and quickly search for lensing pairs among thousands of events, including the sub-threshold candidates.
\end{abstract}


\maketitle

{\bf \em Introduction---}
The first decade of Gravitational Wave (GW) Astronomy has brought massive improvements in our understanding of the universe, e.g., in constraining the formation and population models of stellar-mass compact objects, origin of heavy elements in the universe, various tests of General Relativity, probing the expansion of the universe, etc. Nevertheless, it is still the beginning; a vast set of groundbreaking science is expected from this new window of Astronomy.
One of the most important discoveries expected in the near future is the Gravitational Lensing (GL) of GWs, which can shine light on the distribution of dark matter and constrain other cosmological parameters~\cite{oguriStrongGravitationalLensing2019, liaoStronglyLensedTransient2022}.

Detection of strongly lensed GW events is a highly active area of research~\cite{hannukselaSearchGravitationalLensing2019, mcisaacO1O2SearchStronglyLensed2020, collaborationO3aSearchLensingSignatures2021, collaborationO3SearchGravitationalLensing2023, janquartO3FollowUpAnalysis2023, collaborationO4aSearchGravitationalLensing2025}.
Presently, the most reliable methods employ joint parameter estimation for source and lensing parameters~\cite{janquartFastPreciseMethodology2021, loBayesianStatisticalFramework2023}. However, these methods are computationally intensive and may work only for relatively strong signals, for which reliable parameter estimation is possible. Computationally inexpensive methods, based on direct cross-correlation~\cite{chakrabortyGLANCEGravitationalLensing2024, chakrabortyGLANCEFalseAlarm2025}%
~\footnote{
    While GLANCE is also a direct CC-based method, we introduce an optimal CC estimator here, which makes a huge difference in efficiency, as we find in our study. Also, GLANCE incorporates sky-localization information and requires at least two non-coaligned detectors. However, in this work, we focus on the performance of the optimal estimator, without accounting for sky-localization information that can bring further orders of magnitude improvements in efficiency.
},
phase consistency~\cite{ezquiagaPhazapIdentifyingStrongLensed2023},
machine learning~\cite{goyalRapidIdentificationStrongly2021, magareSLICKStrongLensing2024}, comparison of chi-squares~\cite{gholapStatisticIdentification2025}, and posterior overlap of unlensed parameters~\cite{harisIdentifyingStronglyLensed2018, barsodeFastEfficientBayesian2024} have also been proposed, which either do not use an optimal statistic or are strongly model-dependent. The cross-correlation-based method we propose here, OCCAM, utilizes an optimal statistic with mild model dependence and the noise power spectral density (PSD) for the first time, making it a highly (if not the most) efficient method among computationally inexpensive ones.

{\bf \em Strong Lensing---}
When a GW passes close to a massive lens, it leads to the production of multiple temporally resolved copies of the source GW signal. This regime is known as strong lensing. The properties of the images depend on the lens profile and source lens geometry. For example, if GWs from stellar-mass compact binary coalescences (CBCs) encounter galaxies or galaxy clusters, strong lensing will occur, with a time delay between the images spanning minutes to months for galaxy-scale lenses and weeks to years for cluster-scale lenses. The waveform of each of the lensed images, $h^\text{L}_j$, will have an identical phase evolution, but a change in its amplitude, arrival time, and phase compared to the source signal $h$,
\begin{equation}
    h_j^\text{L}(f) =
    \sqrt{|\mu_j|} \exp{\left[ 2\pi if \Delta t_j - i \sign(f) n_j \pi \right]} h(f),
\end{equation}
where $n_j$ is called the Morse factor for the $j^\text{th}$ image, and takes the value $0,1/2,1$ for Type I, II, and III images, respectively. It is evident from the equation that the magnification $\mu_j$ and the time delay $\Delta t_j$ do not affect the waveform morphology (they are completely degenerate with the luminosity distance and coalescence time), the lensing Morse phase shift $n_j \pi$ can introduce distortions for signals containing higher modes~\cite{daiWaveformsGravitationallyLensed2017, ezquiagaPhaseEffectsStrong2021}. Particularly, this occurs for Type II images, whereas Type I has no phase shift, and Type III only flips the overall sign, which is degenerate with a polarization angle shift by $\pi/2$. The $\sign(f)$ function ensures that the time domain waveform stays real.

With the above definition of strongly lensed waveforms, the relation between a pair of strongly lensed GW strains becomes,
\begin{equation}
    \frac {h_2^\text{L}(f)} {h_1^\text{L}(f)} = \sqrt{|\mu_\text{rel}|} \, \exp{\left[ 2\pi i f \Delta t_\text{rel} - i \sign(f) n_\text{rel} \pi \right]}, \label{eq:lens-pair}
\end{equation}
where $\mu_\text{rel}$ is the relative magnification, defined by the ratio of the two magnifications, while $\Delta t_\text{rel}$ and $n_\text{rel}$ are the differences in time delays and Morse factor between the two images: $j=1$ and $2$.

{\bf \em Methodology---}
We present the methodology in three parts: starting with the formalism of cross-correlation between two data strains from the same or a pair of detectors and the case of strong lensing; we show how we combine the correlations in the case of multiple detectors; finally, we end with how we can apply the method in the case of GW strain data.

{\em Formalism---}
Considering two data strains $\{s_1, s_2\}$ consisting of GW events $\{h_1, h_2\}$ and noises $\{n_1, n_2\}$, i.e., $s_i = h_i + n_i$, $i \in \{1,2\}$, we define an estimator with a filter $Q$,
\begin{equation}
    \hat s(t) =
        \infint df \, \tilde s_1^*(t;f) \tilde s_2(t;f) \tilde Q^*(t;f),
\end{equation}
where, the finite time Fourier transforms of time ($t$) dependent variables, say, $A(t)$, are represented by $\tilde A(t;f)$%
~\footnote{
    As we are dealing with finite time segments $[t-T/2, t+T/2]$, the Fourier pairs are related in the following way,
    \begin{align*}
        \tilde A(t;f) &=
            \int_{t-T/2}^{t+T/2} dt' \, A(t') e^{-2\pi i f t'}, \\
        A(t) &=
            \infint df \tilde A(t;f) e^{2\pi i f t}.
    \end{align*}
}.
Henceforth, we drop the dependency notations (unless explicit mention is necessary) for brevity.
The (expected) signal-to-noise ratio (SNR) of the above estimator is found to be (see Supplemental Material for the derivation),
\begin{equation}
    \rho =
        \frac{\langle(\tilde u, \tilde s_1^* \tilde s_2)_t\rangle}
            {(\tilde u, \tilde u)^{1/2}_t}
        = \frac{(\tilde u, \tilde h_1^* \tilde h_2)_t}
            {(\tilde u, \tilde u)^{1/2}_t} = (\hat u, \tilde h_1^* \tilde h_2)_t,
\end{equation}
with, $\tilde u =\tilde Q \xi$ and,
\begin{align}
    (\tilde A, \tilde B)_t &=
        \infint df \, \frac{\tilde A^* \tilde B}{\xi}\,,
    \label{eq:inner-product} \\
    \xi &=
        \frac 12 |\tilde h_1|^2 S_{n,2}
        + \frac 12 |\tilde h_2|^2 S_{n,1}
        + \frac{T}{4} S_{n,1} S_{n,2},
    \label{eq:modified-psd}
\end{align}
and $S_{n,i}(t;|f|)$ denoting the one-sided noise PSD of the $i^\text{th}$ detector~\footnote{In Ref.~\cite{allenDetectingStochasticBackground1999}, $\xi$ contains only noise PSDs of the detectors, since to search for a stochastic background, the low-signal limit could be assumed, which is not applicable here.}.

In this vector space defined by the inner product (Eq.~\ref{eq:inner-product}), $\hat u = \tilde u / ||\tilde u||$ is a normalized vector with norm defined as $||\tilde u|| = (\tilde u, \tilde u)^{1/2}$. In that sense, one can argue that the SNR, $\rho$, would be maximized when $\tilde u$ is parallel to $\tilde h_1^* \tilde h_2$, i.e.,
\begin{align}
    \tilde u &\propto \tilde h_1^* \tilde h_2, \\
    \tilde Q &\propto \tilde h_1^* \tilde h_2 / \xi,
\end{align}
leading to the cross-correlation (CC) SNR estimator,
\begin{equation}
    \rho_\text{CC} =
        \frac{(\tilde h_1^* \tilde h_2, \tilde s_1^* \tilde s_2)_t}
            {(\tilde h_1^* \tilde h_2, \tilde h_1^* \tilde h_2)_t^{1/2}}\,,
    \label{eq:cc-SNR}
\end{equation}
which corresponds to the expected optimal SNR,
\begin{equation}
    \rho_\text{opt} = \langle \rho_\text{CC} \rangle =
        (\tilde h_1^* \tilde h_2, \tilde h_1^* \tilde h_2)_t^{1/2}.
    \label{eq:optimal-SNR}
\end{equation}

Eq.~\eqref{eq:cc-SNR} is general in the sense that we have not assumed anything about the relations between the two GW events. Also, it allows the two signals to be from the same or different detectors.

Since we are particularly interested in strongly lensed signals here, we introduce the relation between the GW events through Eq.~\eqref{eq:lens-pair}. We ignore the time delay factor as it is completely degenerate with the time of coalescence, but we keep the relative magnification $\mu_\text{rel}$ and Morse phase difference $m_\phi = n_\text{rel} \pi$. The factor $\sign(f)$ becomes $1$ since we keep frequencies always positive. This leads to the relation between $h_1$ and $h_2$ as,
\begin{equation}
    \tilde h_2 =
        \tilde h_1 \sqrt{|\mu_\text{rel}|} e^{im_\phi}.
    \label{eq:strong-lensing}
\end{equation}
In such a scenario, the Eqs.~(\ref{eq:modified-psd},\ref{eq:optimal-SNR},\ref{eq:cc-SNR}) get modified as,
\begin{align}
    \xi =
    \frac 14 \left[
        2 |\tilde h_1|^2 (|\mu_\text{rel}| S_{n,1} + S_{n,2}) + T \, S_{n,1} S_{n,2}
    \right],
    \label{eq:modified-psd-sl} \\
    \rho_\text{opt} =
        \sqrt{|\mu_\text{rel}|}(|\tilde h_1|^2, |\tilde h_1|^2)_t^{1/2},
    \label{eq:optiml-SNR-sl} \\
    \rho_\text{CC} =
        \frac{(|\tilde h_1|^2, \tilde s_1^* \tilde s_2)_t}
        {(|\tilde h_1|^2, |\tilde h_1|^2)_t^{1/2}}
        e^{-im_\phi}.
    \label{eq:cc-SNR-sl}
\end{align}
We maximize over the phase by taking the absolute value of Eq.~\eqref{eq:cc-SNR-sl}, which eliminates the $e^{-im_\phi}$ factor as it is simply an overall phase shift,
\begin{align}
    \rho_\text{CC}^{\max} \underset{\max \{\phi\}}{=}
        \frac{\abs \left( (|\tilde h_1|^2, \tilde s_1^* \tilde s_2)_t \right)}
        {(|\tilde h_1|^2, |\tilde h_1|^2)_t^{1/2}}\,.
    \label{eq:cc-SNR-maximized}
\end{align}
Hence, Eq.~\eqref{eq:cc-SNR-maximized} finalizes our cross-correlation SNR estimator, where the inner product $(\,,)$ is defined by Eq.~\eqref{eq:inner-product} and $\xi$ by Eq.~\eqref{eq:modified-psd-sl}, which requires the following quantities:
\begin{itemize}
    \item the duration, $T$, of the time segment used for the analysis,

    \item relative magnification, $\mu_\text{rel}$, estimated from the matched-filter (MF) SNR~\cite{allenFINDCHIRPAlgorithmDetection2012}, $\rho_\text{MF}^{(i)}$, for the $i^\text{th}$ event%
    ~\footnote{
        MF SNR is defined as, $\rho_\text{MF} = (\tilde h, \tilde s)_\text{MF}/(\tilde h, \tilde h)_\text{MF}^{1/2}$, and optimal MF SNR as, $\rho_\text{MF}^\text{opt} = (\tilde h, \tilde h)_\text{MF}$, where the MF inner product $(\,,)_\text{MF}$ is defined in Eq.~\ref{eq:mf-inner-product}.
    }, $|\mu_\text{rel}| = (\rho_\text{MF}^{(2)}/\rho_\text{MF}^{(1)})^2$,

    \item $|\tilde h_1(t;f)|^2$, for which we use best-fit template from the MF based search%
    ~\footnote{
        Note that the method is primarily a model-independent search. The weight factor used to optimize the SNR depends (weakly) on the power spectrum, not on the explicit phase evolution. Also, the antenna pattern functions, being essentially constant throughout the duration of the events, introduce constant phase shifts to the signals without altering the signal morphology, which does not affect our CC statistic.
    }.
\end{itemize}

As we are dealing with signals of different strengths, the estimator's value, being proportional to SNR, would also vary. In such a scenario, we normalize the output to
[0,1] (with noise fluctuations), where values near $0$ indicate weak correlation (unlensed) and values near $1$ indicate strong correlation (lensed). Setting a threshold to determine the binary classification would be unique and fixed in this scenario. The normalization is done by dividing by the expected SNR of the output, which is the optimal SNR, leading to the normalized CC SNR as,
\begin{align}
    \hat \rho_\text{CC}^{\max} =
        \frac{\rho_\text{CC}^{\max}}{\rho_\text{opt}}\,.
    \label{eq:cc-SNR-max-norm}
\end{align}
For future references, we adopt the notation $\rho_\text{CC}(s_i,s_j) \equiv \sfrac{(|\tilde h_1|^2, \tilde s_i^* \tilde s_j)_t e^{-im_\phi}}{(|\tilde h_1|^2, |\tilde h_1|^2)_t^{0.5}}$, and similarly for other variables, e.g., $\rho_\text{CC}^{\max}(s_i,s_j), \hat \rho_\text{CC}^{\max}(s_i,s_j)$, to denote the cross-correlation between $s_i$ and $s_j$. Furthermore, we also define the numerator of $\rho_\text{CC}^{\max}$ by,
\begin{equation}
    \rho^{\max} (s_i, s_j) \equiv
        \abs \left((|\tilde h_1|^2, \tilde s_i^* \tilde s_j)_t \right).
\end{equation}
Think of $\rho$ as the raw correlation, and when divided by $\rho_\text{opt}$, we get the CC SNR $\rho_\text{CC}$, and finally $\hat \rho_\text{CC}$ after normalization (with `max' standing for the maximization over phase).
In the text that follows, we will drop the `max' in notation, as we will always take the SNR maximized over phase%
~\footnote{
    Extrinsic parameters do not affect our statistic, as they introduce only global amplitude and phase shifts, which are already accounted for by the normalized, phase-optimized CC statistic.
}.

{\em Multiple detectors---}
The formalism, by definition, can be easily extended to multiple detectors. If we consider $s_i^{(D_j)}$ to be the $i^\text{th}$ data strain in $j^\text{th}$ detector that has PSD $S_n^{(D_j)}$, then to cross-correlate between the data strain $s_i^{(D_k)}$ and $s_j^{(D_l)}$, what we want to calculate is $\hat \rho_\text{CC}(s_i^{(D_k)}, s_j^{(D_l)}) = \sfrac{\rho}{\rho^2_\text{opt}} (s_i^{(D_k)}, s_j^{D_l})$ (see Eq.~\eqref{eq:cc-SNR-max-norm}), where we put $S_{n,1} = S_n^{(D_k)}$ and $S_{n,2} = S_n^{(D_l)}$ in the modified PSD Eq.~\eqref{eq:modified-psd-sl}.
To combine the SNRs, we create a naive estimator, which may not be optimal, by taking an average over the raw correlations $\rho$ and the squared optimal SNRs $\rho^2_\text{opt}$, and then taking the ratio, i.e.,
\begin{equation}
    \hat \rho_\text{multi-det} =
        \frac {\sum_{D_i, D_j \in \{\text{H1}, \text{L1}, ...\}} \rho(s_1^{D_i}, s_2^{D_j})}
            {\sum_{D_i, D_j \in \{\text{H1}, \text{L1}, ...\}} \rho_\text{opt}^2(D_i, D_j)}\,.
    \label{eq:multi-det}
\end{equation}
Note that the detector dependencies of $\rho_\text{opt}$ are coming from the modified PSD equation, which defines the inner product. The square on $\rho_\text{opt}$ is coming from the fact that we divide $\rho$ twice to get to the normalized SNR.

We do not perform self-correlations i.e., $\rho(s_i^{(D_j)}, s_i^{(D_k)})$, since we are interested in searching for the correlations between $s_1$ and $s_2$. This indicates that, for $N$ detectors, we are computing $N^2$ cross-correlations.

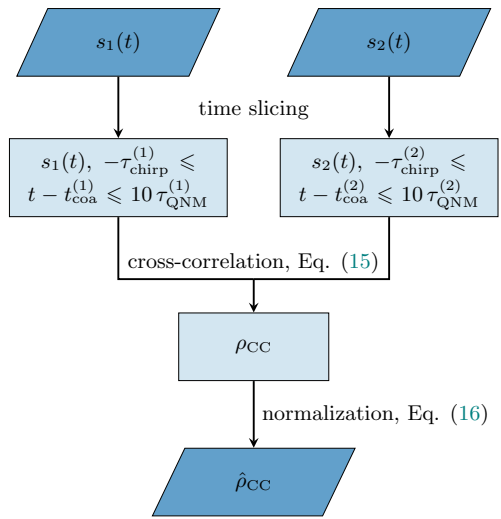
\begin{figure}[tb]
    \scalebox{0.9}{
        \begin{tikzpicture}[node distance=2cm]

    \node (ts1) [io] {$s_1(t)$};
    \node (ts2) [io, right of=ts1, xshift=2cm] {$s_2(t)$};
    \node (ts1_chunk) [process, below of=ts1, yshift=-0cm] {$s_1(t), ~ -\tau_\text{chirp}^{(1)} \leqslant t - t_\text{coa}^{(1)} \leqslant 10 \, \tau_\text{QNM}^{(1)}$};
    \node (ts2_2) [process, below of=ts2, yshift=-0cm] {$s_2(t), ~ -\tau_\text{chirp}^{(2)} \leqslant t - t_\text{coa}^{(2)} \leqslant 10 \, \tau_\text{QNM}^{(2)}$};

    \node (CC) [process, text width=2cm] at (+2, -4.5) {$\rho_\text{CC}$};
    \node (out) [io, below of=CC, yshift=-0cm] {$\hat \rho_\text{CC}$};

    \node[draw=none] at (2, -1) {time slicing};
    \node[draw=none] at (+2, -3.25) {cross-correlation, Eq. \eqref{eq:cc-SNR-maximized}};
    \node[draw=none] at (+3.8, -5.5) {normalization, Eq. \eqref{eq:cc-SNR-max-norm}};

    \draw [arrow] (ts1) -- (ts1_chunk);
    \draw [arrow] (ts2) -- (ts2_2);
    \draw [line] (ts1_chunk) |- (2, -3.5);
    \draw [line] (ts2_2) |- (2, -3.5);
    \draw [arrow] (2, -3.5) -- (CC);
    \draw [arrow] (CC) -- (out);

\end{tikzpicture}
    }
    \caption{\label{fig:CC-flowchart}%
    Flowchart illustrating the application of the cross-correlation method for identifying strongly lensed GW candidates. Before cross-correlating $s_1$ and $s_2$ according to the formalism developed in this paper, we perform \textit{time slicing} on $s_1$ and $s_2$, as shown in Fig.~\ref{fig:cross-correlation}. After performing cross-correlation, we normalize the CC output, $\rho_\text{CC}$ using Eq.~\eqref{eq:cc-SNR-max-norm} to obtain the final output, $\hat \rho_\text{CC}$. We take this output to be the result of our method.}
\end{figure}

\begin{figure*}[htb]
    \includegraphics[width=0.85\linewidth]{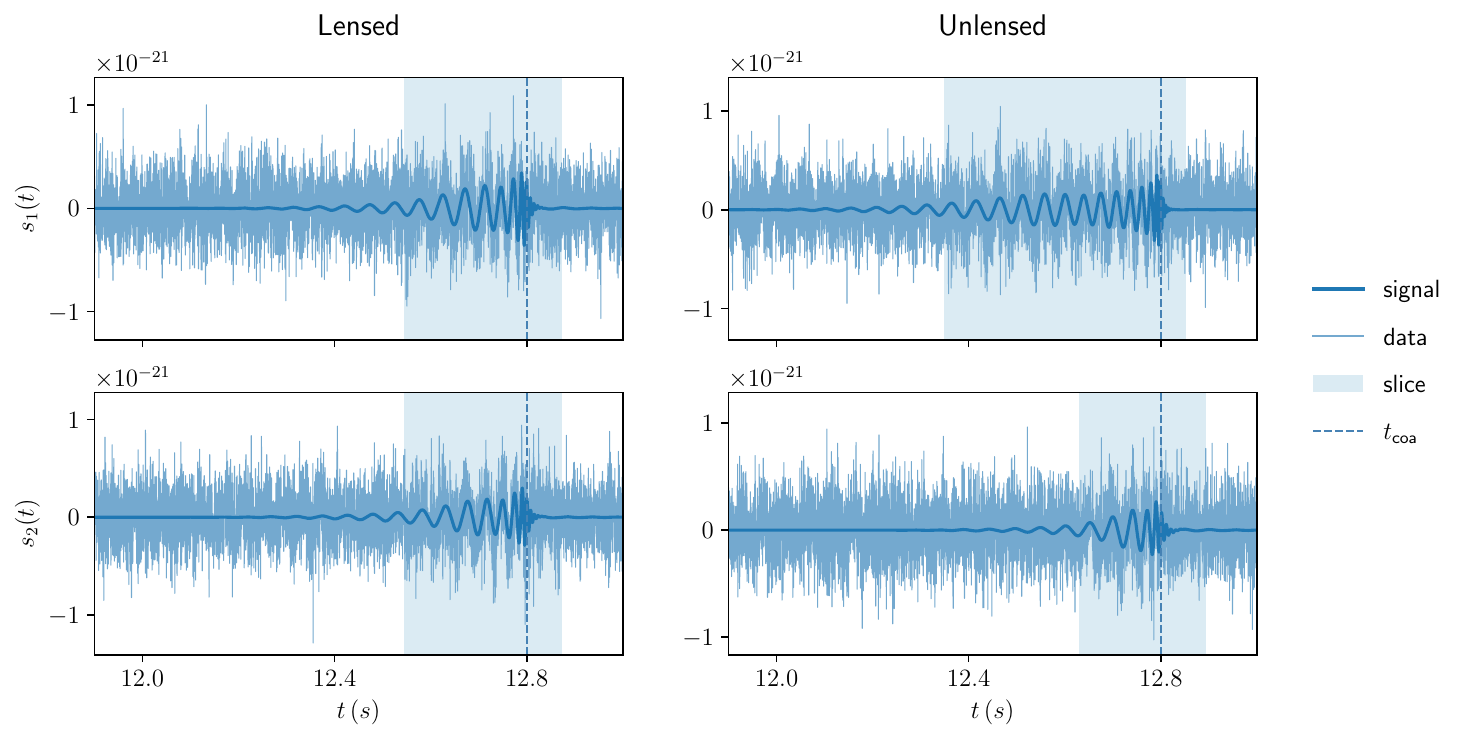}
    \caption{\label{fig:cross-correlation}%
    Figure displaying chopping width in \textit{time-slicing} operation. The two subfigures on the left(right) designate lensed(unlensed) scenario, while the top(bottom) row shows $s_1(s_2)$. The data (light blue) contains noise and the signal (blue). We chop the data such that $t-t_\text{coa} \in [-\tau_\text{chirp}, 10\tau_\text{QNM}]$, which is depicted by the shaded area (light blue). We take these chopped $s_1$ and $s_2$ and perform the \textit{cross-correlation} (Eq.~\eqref{eq:cc-SNR-maximized}), after making sure that the time of coalescences, $t_\text{coa}$ (shown as blue dashed line), of the events are aligned. For the above example, the CC statistic, $\hat \rho_\text{CC}$, was $0.97$ for lensed and $0.17$ for unlensed.}
\end{figure*}

{\em Application---}
In Fig.~\ref{fig:CC-flowchart}, we show a flowchart of the procedure for applying the formalism developed in this paper.
In the following paragraphs, we describe the \textit{time slicing} operation.

Since both the data strains have noise present, to make sure that the CC output comes predominantly from the signal rather than some random noise fluctuations, we slice both $s_1$ and $s_2$ around the event in such a way that the segment of data contains mostly the event, as shown by the light blue block in Fig.~\ref{fig:cross-correlation}.
This segment is decided from the time of coalescence $t_\text{coa}$, by taking the chirp time $\tau_\text{chirp}$ as a representative of the inspiral, and ten times the quasi-normal mode (QNM) damping time%
~\footnote{
    Since the damping time is the time it takes for the amplitude to drop by a factor of e, we used $10$ times the damping time to ensure the amplitude falls below $0.01\%$.
}
of 22 mode $\tau_\text{QNM}$ for the ringdown%
~\footnote{
    Only a representative signal duration is required. Thus, using the 22-mode QNM damping time, which is the largest, is a good approximation. Similarly, excluding the merger time in the inspiral is justified, as it is much shorter.
}.
The expressions for $\tau_\text{chirp}$~\cite{cokelaerGravitationalWavesInspiralling2007} and $\tau_\text{QNM}$~\cite{bertiQuasinormalModesBlack2009} are as follows%
~\footnote{
    These expressions require information about the event (model dependency), but we do not require explicit phase information; only the event's masses are needed, which can be determined from the MF best-fit template parameters.
}%
,
\begin{align}
    \tau_\text{chirp}
        &= \frac{5}{256 \pi \eta f_\text{low}} (\pi M_\text{tot} f_\text{low})^{5/3} \, \text{s}, \label{eq:chirp-time}
        \\
    \tau_\text{QNM}
        &= 0.5537 \times \left( \frac {M}{10 M_\odot} \right) \, \text{ms}. \label{eq:QNM-damp-time}
\end{align}
If we do not consider the \textit{time slicing}, the method's efficiency decreases; see End Matter.

After \textit{time slicing}, we have $s_i(t)$ with $t$ ranging from $t_\text{coa}^{(i)} - \tau_\text{chirp}^{(i)}$ to $t_\text{coa}^{(i)} + 10\tau_\text{QNM}^{(i)}$ for $i \in \{1,2\}$. We make sure that the coalescence times for the two sliced data strains ($t_\text{coa}^{(1,2)}$) are aligned properly, and then we perform \textit{cross-correlation} among them by invoking Eq.~\eqref{eq:cc-SNR-maximized}, where $|\tilde h_1|^2$ is set to be the matched-filter best fit template of the higher $\rho_\text{MF}$ among the two. The output value is then normalized using Eq.~\eqref{eq:cc-SNR-max-norm}. The final normalized output lies in the range $[0,1]$ (with noise fluctuations). If the final output exceeds a certain threshold, then we declare the pair of GW events to be (potentially) lensed; otherwise, it is considered unlensed. The threshold can be determined by conducting a noise background study.

{\bf \em Background Analysis \& Result---}
To test the method's efficiency and set the threshold, we simulate lensed and unlensed events, as described in Ref.~\cite{moreImprovedStatisticIdentify2022}.
Not all events are detectable in the GW detectors, as they are preferentially sensitive to louder signals. We select only those events with an optimal MF SNR, $\rho_\text{MF}^\text{opt} \geqslant 10$ in the LIGO Hanford detector, assuming the zero-detuned-high-power PSD of Advanced LIGO at design sensitivity~\cite{aLIGOZeroDetHighPowerPSD2009} with a lower frequency cutoff at $20$ Hz and \texttt{IMRPhenomD}~\cite{husaPhenomD2016, khanPhenomD2016} waveform approximant%
~\footnote{
 IMRPhenomD should be adequate since we are using non-spinning injections.
}.
With this selection criterion, we have $1146$ pairs of Type I \& II lensed events, and $1703$ unlensed events. This leads to a background of $^{1703}C_2 \sim 1.5 \times 10^6$ pairs, where we combine the unlensed events pairwise, making $1$ lensed event in $\sim$$1000$ unlensed events. Refer to the End Matter for more details on dataset generation.

To simulate a realistic scenario, we first perform a matched-filter search using a template bank%
~\footnote{
    We generate a non-spinning geometric template bank with a minimum match of 0.97, ensuring that all injections (without noise) are recovered with a match (Eq.~\eqref{eq:match}) value greater than 0.97.
},
and take the best-fit trigger's characteristics to set $t_\text{coa}$, $\tau_\text{chirp}, \tau_\text{QNM}, \mu_\text{rel}$ and $|\tilde h_1(f)|^2$. Following the flowchart in Fig.~\ref{fig:CC-flowchart}, we chop the two data strains and perform our method of cross-correlation. Since we are using template-based MF search to identify triggers, it may happen that the inferred $t_\text{coa}$ differs slightly from the actual time of coalescence. As our statistic $\hat \rho_\text{CC}$ correlates $s_1$ and $s_2$ with times aligned at $t_\text{coa}$, one might argue that wrong identification of $t_\text{coa}$ will lead to wrong inference.
We find that allowing for a time lag in the cross-correlation does not improve performance; the best performance is obtained when it is omitted. This is because the noise contribution overwhelmingly dominates any benefits from correctly identifying the time of coalescence (See Supplemental Material for details).

\begin{figure}[htb]
    \includegraphics[width=0.9\linewidth]{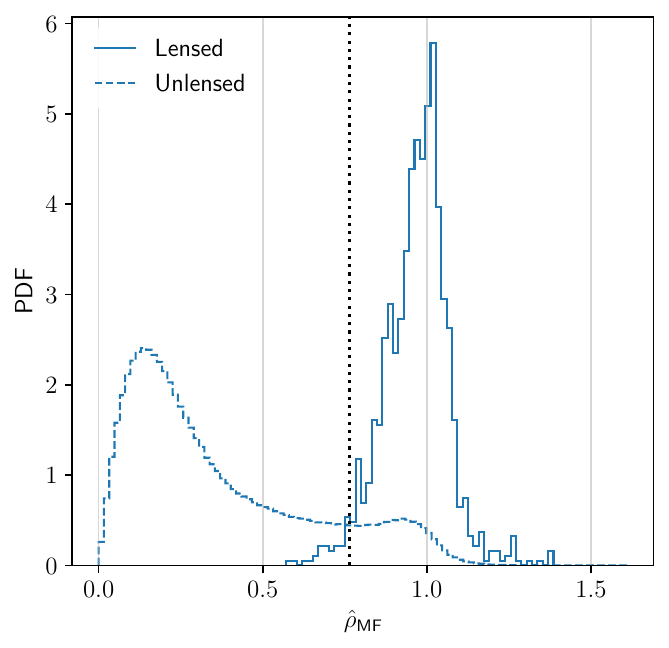}\\
    \includegraphics[width=0.9\linewidth]{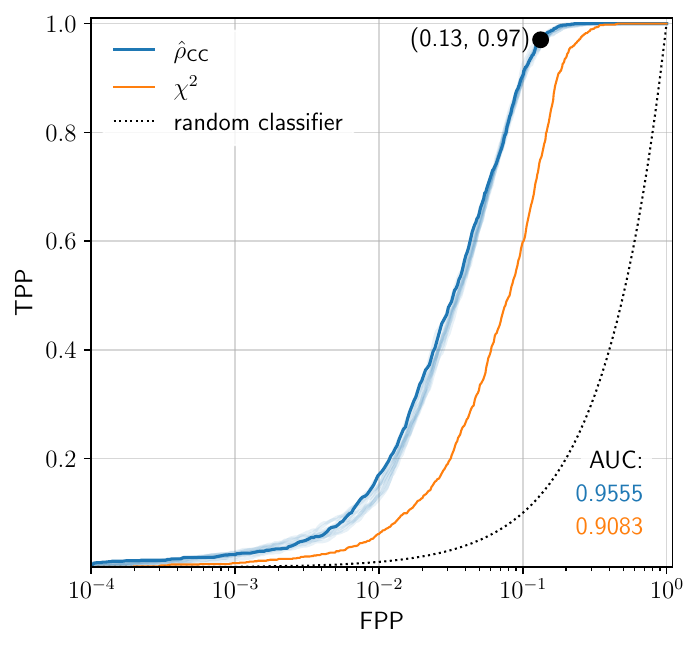}
    \caption{\label{fig:result-H1}%
    Histogram (top) and ROC with AUC (bottom) shown for cross-correlation analysis performed on simulated lensed and unlensed events based on a realistic astrophysical population with $\rho_\text{MF}^\text{opt} \geqslant 10$. $\hat \rho_\text{CC}$ (blue) corresponds to the method described in the text of background analysis, while we compare our method with that of the $\chi^2$ based analysis~\cite{gholapStatisticIdentification2025} (orange). Histogram is shown only for $\hat \rho_\text{CC}$. The light blue ROC curves correspond to several different runs with the same injections but in different noise realizations. We also indicate a threshold (black dotted vertical line) in the histogram, and the corresponding (FPP, TPP) coordinate (black dot) in the ROC plot.
    }
\end{figure}

We show the distribution of our $\hat \rho_\text{CC}$ statistic for lensed and unlensed events in Fig.~\ref{fig:result-H1}. We observe a small overlap between the lensed and unlensed distributions, which can be attributed to accidentally having two unlensed events with similar CBC parameters, or the CC statistic producing a lower output value in the lensed case due to noise uncertainties. Particularly, the second peak of the unlensed distribution near $\hat \rho_\text{CC}=1$ occurred because of the similarity between a few unlensed pairs in the dataset we used. (See the End Matter for additional visualizations of the dataset, which also reveal that the optimal statistic performs significantly better.) A good classifier should minimize this overlap. With increased SNR, this overlap is expected to reduce, leading to better discrimination (see End Matter).

To test the performance of a binary classifier like ours (lensed vs unlensed), we plot the receiver-operator-characteristics (ROC) curve, which denotes the efficiency of correctly identifying true positive (lensed) events (TPP) as a function of false positive probability (FPP, probability of falsely classifying unlensed events to be lensed). Higher TPP for a small FPP is preferred for a better model. We also report the area under the ROC curve (AUC) as a quantitative metric; a higher AUC indicates better discrimination.
Fig.~\ref{fig:result-H1} %
\footnote{
    The ROC obtained using injection parameters is not significantly different from that obtained using trigger parameters; only the latter is shown here (see Supplemental Material for more details).
}
shows this ROC with AUC, where we also compare our method's performance (using the same set of injections) with that of Ref.~\cite{gholapStatisticIdentification2025}, keeping note that they had concluded their method's performance to be at par with the other current methods~\cite{harisIdentifyingStronglyLensed2018,goyalRapidIdentificationStrongly2021}%
~\footnote{
    The ROC curves for GLANCE were not available in the corresponding publications~\cite{chakrabortyGLANCEFalseAlarm2025}
}.
We show an operating point (threshold on $\hat \rho_\text{CC}$) corresponding to an FPP of $10\%$.
We find that our method ($\hat \rho_\text{CC}$) shows a significant improvement over their method ($\chi^2$).

\begin{figure}[htb]
    \includegraphics[width=0.9\linewidth]{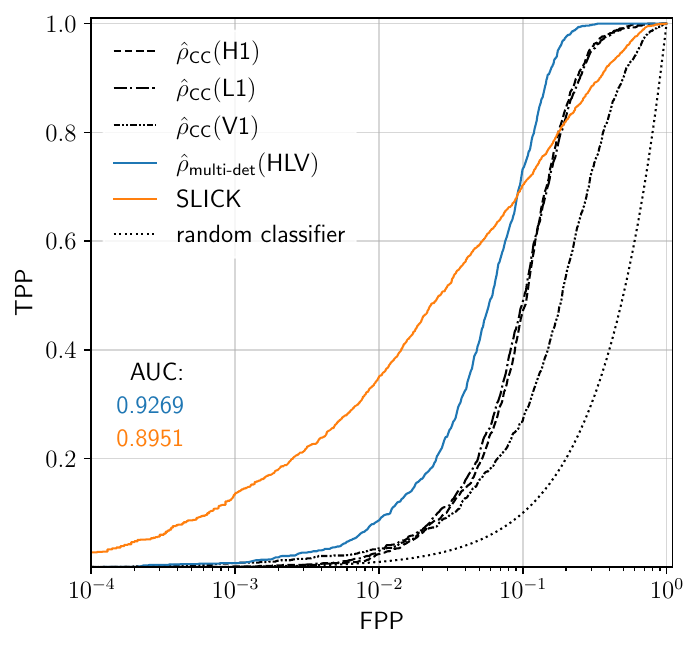}
    \caption{\label{fig:result-netw}%
    ROC shown after performing multi-detector cross-correlation for a network of HLV detectors for network $\rho_\text{MF}^\text{opt} \geqslant 8$ and individual $\rho_\text{MF}^\text{opt} \geqslant 4$. As desired, the ROC for the network (blue) is much better compared to individual detector responses (black). We have also added the ML-based SLICK pipeline~\cite{magareSLICKStrongLensing2024} ROC curve (orange) for reference.
    }
\end{figure}

Finally, we share the results of the multi-detector analysis. We take design sensitivities of all the respective detectors - Hanford, Livingston, and Virgo~\cite{aLIGOZeroDetHighPowerPSD2009, AdvVirgoPSD2012}.
Given these sensitivities, we select events with an optimal MF SNR ($\rho_\text{MF}^\text{opt}$) above the network threshold of 8 and the individual detector threshold of 4. The sample size is kept the same as before. We compute $\hat \rho_\text{multi-det}$ for the HLV network using  Eq.~\eqref{eq:multi-det}, which has $3^2$ correlation terms. Note that we did not perform an MF search for this analysis; instead, we took the injection values. We only show ROCs for the individual detectors and the combined network response in Fig.~\ref{fig:result-netw}.
As desired, the ROC of the network of detectors performs significantly better than that of the individual detectors. We have also included the ROC curve for the machine learning (ML) based SLICK pipeline~\cite{magareSLICKStrongLensing2024} for reference, although the set of injections used here may differ.%

{\bf \em Conclusion and Discussion---}
The optimal cross-correlation-based detection algorithm we present here for detecting strong GW lensing pairs, OCCAM, performs highly efficiently. We demonstrate its capabilities rigorously using simulated data, with both Type I \& II lensed events, and ROC curves, and by comparing with similar methods (where such comparisons were possible). Apart from its ability to quickly identify strongly lensed pairs with a relatively high true detection probability at a given false positive probability, this method requires an insignificant amount of computation time, without requiring any derived data products that can introduce additional uncertainties, and is capable of working with single detector events, making it an excellent method for detecting lensing candidates. While we have tested our method for quasi-circular BBH and relatively high-SNR events, it has the potential to work for any CBC signals (including neutron stars), with higher-order effects.

We are yet to account for the sky-localization information, which is usually available from the low-latency analyses. This will likely reduce the false positive probability, perhaps by one or two orders of magnitude~\cite{harisIdentifyingStronglyLensed2018}, making it viable to efficiently look for potential lensing pairs among thousands of CBC detection candidates~\cite{collaborationGWTC4aCandidates2025}.

{\bf \em Acknowledgments---}
We thank Sudhir Gholap, Bhooshan Gadre, Srasthi Goyal, Shasvath Kapadia, and Parameswaran Ajith for valuable discussions. We sincerely thank the authors of Ref.~\cite{gholapStatisticIdentification2025} for sharing their scripts with us for performing the comparison. We also acknowledge the use of the IUCAA LDG cluster Sarathi for carrying out the computational and numerical work presented in this study. This research was made possible through the financial support of the University Grants Commission, Government of India. We acknowledge the use of the following \textsc{Python} packages: Numpy~\cite{numpy2020}, Scipy~\cite{scipy2020}, Pandas~\cite{pandas2020}, Matplotlib~\cite{matplotlib2007}, Astropy~\cite{astropy2022}, PyCBC~\cite{pycbc2024}.

%

\section{End Matter}

{\bf \em Simulated data generation---}
We follow Ref.~\cite{moreImprovedStatisticIdentify2022} to generate lensed and unlensed populations of binary black holes (BBHs) and refer the reader to it for additional details.
We assume a power law plus a peak BBH mass model, with the merger rate density model following Ref.~\cite{oguriEffectGravitationalLensing2018}. We use the \texttt{IMRPhenomPv2}~\cite{hannamPhenomPv22014, khanPhenomPv22016} waveform approximant%
~\footnote{
    Note that in our analysis, we use \texttt{IMRPhenomD}, which does not cause any inconsistencies since \texttt{IMRPhenomPv2} is based on \texttt{IMRPhenomD} and the sample space does not contain any spinning binaries, resulting in both approximants producing the same waveforms.
},
implemented in \textsc{LALSuite}~\cite{lalsuite2018}, to generate the BBH signal injected in the aLIGO Gaussian noise.
For the lens population, we consider massive early-type galaxies as the lens galaxies, whose density profiles follow a Singular Isothermal Ellipsoid (SIE) lens model~\cite{kormannIsothermalEllipticalGravitational1994}. We have both doubly and quadruply (quad) imaged lenses in the sample, which includes Type I and II images~\cite{ezquiagaPhaseEffectsStrong2021, daiWaveformsGravitationallyLensed2017}.
All events, where the faintest one has a network $\rho_\text{MF}^\text{opt} \geqslant 8$, are included in the full sample, leading to the generation of $\sim$$17000$ lensed pairs and $\sim$$13000$ unlensed events. The sky location parameters, inclination, and polarization are varied in our dataset.

We impose certain conditions based on our requirements --- we choose events with $\rho_\text{MF}^\text{opt} \geqslant 10$ using the Hanford detector in \texttt{aLIGOZeroDetHighPower} PSD. This led to events with a minimum mass of $10 \, \text{M}_\odot$, whereas we constrain the maximum mass to $100 \, \text{M}_\odot$ to have a reasonable number of cycles with the lower cutoff frequency ($20 \, \text{Hz}$) we chose. This resulted in $1703$ unlensed events and $1146$ lensed pairs. Note that we have selected the brightest pairs (the second and third images) from the quad to increase the lensed sample size. This has resulted in a Morse phase difference of $\pi/2$ between the lensed events in any pair. For the multi-detector analysis, we keep the same sample size and impose a network $\rho_\text{MF}^\text{opt} \geqslant 8$ in the HLV network%
~\footnote{
    Note that in the analysis, we use slightly different (O4) sensitivities; see the main text for details.
}
while keeping the same upper mass cut.

\begin{figure}[htb]
    \includegraphics[width=0.9\linewidth]{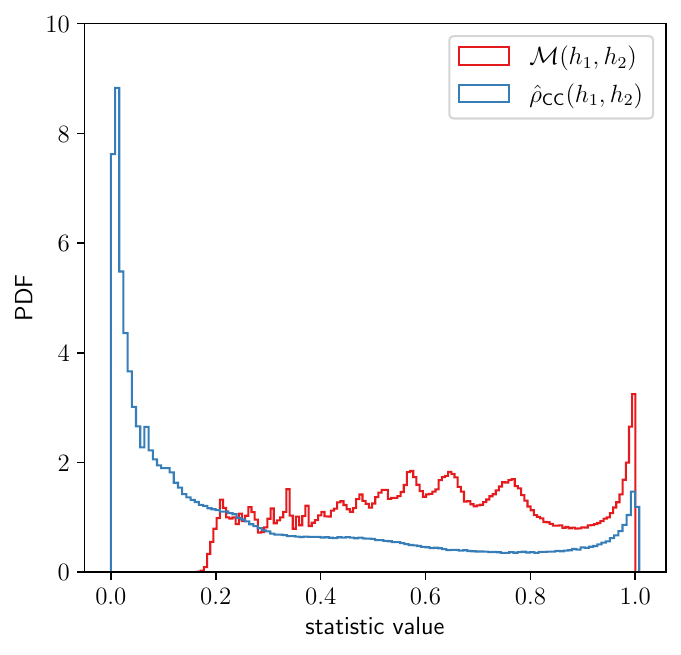}
    \caption{\label{fig:unlens-cc-mat}%
    The distribution of $\mathcal{M}$ and $\hat \rho_\text{CC}$ values for all unlensed pairs. This clearly shows that the optimal statistic outperforms the simple match by significantly reducing false positives. This partly explains why our method can outperform a direct cross-correlation without an optimized estimator.}
\end{figure}

{\bf \em Simulated data inspection---}
The set of injections used for the analysis in Fig.~\ref{fig:result-H1} has a Hanford detector optimal MF SNR threshold of 10, with Advanced LIGO design sensitivity.
We use the same dataset to investigate the reason behind the second peak near $\hat \rho_\text{CC}=1$ for the unlensed distribution shown in Fig.~\ref{fig:result-H1}. To inspect whether that is coming from inherent similarity in the astrophysically motivated unlensed population, we compute the match between the waveforms generated by every unlensed pair, where match denotes the similarity between the two waveforms, defined by
\begin{equation}
    \mathcal{M}(h_1, h_2)
    \equiv (\hat h_1, \hat h_2)_\text{MF},
    \label{eq:match}
\end{equation}
where the MF inner product is defined as,
\begin{equation}
    (\tilde A, \tilde B)_\text{MF}
    \equiv 4 \int_{0}^{\infty} df  \, \frac{\tilde A^*(f) \tilde B(f)}{S_n(f)},
    \label{eq:mf-inner-product}
\end{equation}
and $h_i$ is normalized by its norm $||h_i||=(\tilde h_i, \tilde h_i)_\text{MF}^{1/2}$, $\hat h_i = h_i/||h_i||$.
We also compute our CC statistic without the noise in the data to compare with the match values, i.e.,
\begin{equation}
    \hat \rho_\text{CC}(h_1, h_2)
    = \frac{\abs \left( (|\tilde h_1|^2, \tilde h_1^* \tilde h_2)_t \right) }{(|\tilde h_1|^2, |\tilde h_1|^2)}.
\end{equation}

Fig.~\ref{fig:unlens-cc-mat} displays the histogram of the match values as well as our CC statistic $\hat \rho_\text{CC}$ computed for all unlensed pairs. We find that the lowest match is close to $0.2$, while there is a peak at $1$, indicating that a significant fraction of unlensed pairs with similar parameters, i.e., false positives. This explains why we saw the small second peak in the unlensed distribution of CC statistic in Fig.~\ref{fig:result-H1}. We also observe that our method improves upon it: most event pairs yield a correlation close to $0$, unlike before, while the number of pairs with unit correlation is reduced, thereby reducing false positives. This indicates that our method is optimal for finding correlations between strong lenses compared to any other method that relies on (in some way or other) computing a direct match between pairs of GW signals. We also checked the distribution of $\hat \rho_\text{CC}$ values for the lensed injections without noise (not shown here), and found that all the values lie tightly around the value of $1$ with a standard deviation of $6 \times 10^{-4}$, compared to $0.13$ with the inclusion of noise.
This indicates that all the spread that was seen in Fig.~\ref{fig:result-H1} was due solely to noise.

\begin{figure}[htb]
    \includegraphics[width=0.9\linewidth]{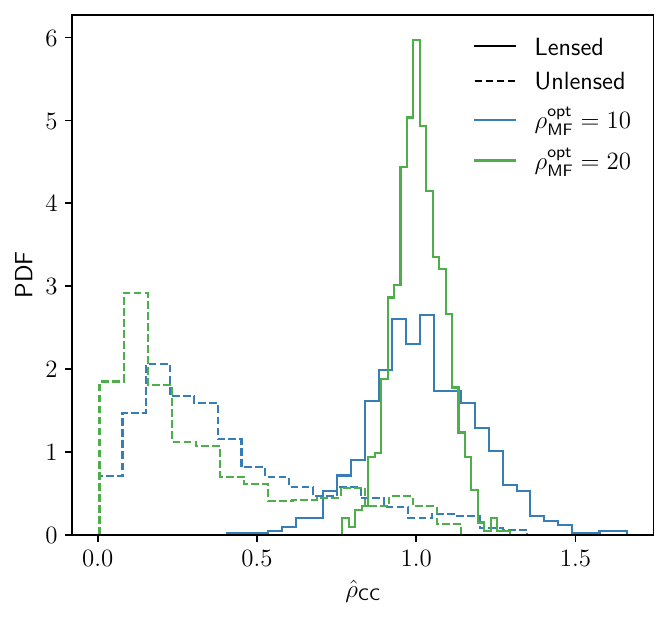}
    \caption{\label{fig:ROC-SNR-variation-hist}%
    Histogram for $\hat \rho_\text{CC}$ statistic of lensed(solid) and unlensed(dashed) events for fixed SNR study. The two colors represent all events being injected with optimal SNR of 10 (blue) and 20 (green). We see that the overlap between lensed and unlensed distributions reduces as SNR increases.}
\end{figure}

\begin{figure}[htb]
    \includegraphics[width=0.9\linewidth]{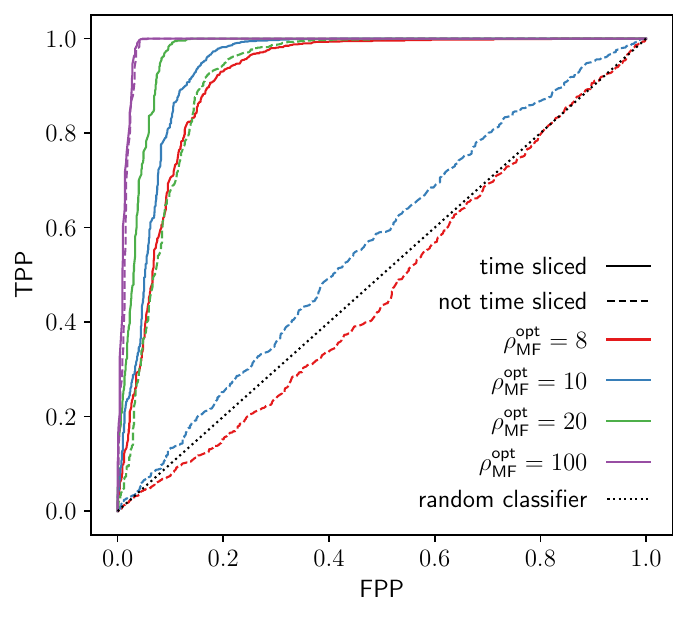}
    \caption{\label{fig:ROC-SNR-variation}%
    ROCs shown for fixed SNR study with and without time slicing. Colors represent different SNR values, while the linestyle denotes whether time slicing was performed(solid) or not(dashed). We see significant improvement in the ROCs after time slicing.}
\end{figure}

{\bf \em Factors affecting efficiency---}
To enhance the method's efficiency in distinguishing between lensed and unlensed GW events, we have conducted several small-scale studies, which are discussed here. In these small-scale studies, we analyzed $1000$ pairs of lensed and unlensed events using our cross-correlation method. First, we have verified the expected variation in the efficiency of such a discriminator with the event SNR. To do so, we have injected all $1000$ events with the same optimal SNR and performed our analysis; we refer to this as the \textit{fixed SNR study}. Fig.~\ref{fig:ROC-SNR-variation-hist} shows the distribution of the cross-correlation statistic $\hat \rho_\text{CC}$ for lensed and unlensed events separately, as a function of the injected SNR. As the signal strength increases, we observe that the variance of the lensed distribution, centered at $1$, decreases. This leads to lower overlap between the lensed and unlensed distributions, thereby improving discrimination performance, as seen in Fig.~\ref{fig:ROC-SNR-variation}.

We had to chop the data strains (around GW events) to reduce the contribution of noise in the CC output. We show, in Fig.~\ref{fig:ROC-SNR-variation}, how this chopping improves the ROC. We have performed the fixed SNR study with and without the \textit{time slicing} step. We observe a significant improvement in the efficiency of discriminating between lensed and unlensed events when \textit{time slicing} is performed, compared to when not. We notice that the ROC for the lowest SNR, when the full data width is considered, performs even worse than a random classifier, which is not unexpected since this scenario is dominated by noise. We also note that, as SNR increases to a really high value, the requirement for \textit{time slicing} decreases. This is expected since \textit{time slicing} was performed to reduce noise contribution to the CC output, and in high SNR, noise contribution is already minimal.

Since our method uses cross-correlation between two data strains that include noise, one can expect that as we increase the sample rate or the signal duration, the performance will degrade. The performance indeed degrades as the sample rate increases, but it remains unaffected if the duration is altered. As we perform the \textit{time slicing} operation, we always fix the data width, limiting the number of noise samples. Thus, with a change in duration, as long as the lower frequency cutoff stays the same (see Eq.~\eqref{eq:chirp-time}), the analysis will stay unaffected. However, if the sample rate is increased, the number of noise samples within the sliced data will also increase, resulting in a degradation in performance, as we do not expect the signal to have any power in larger frequencies.

\begin{figure}[htb]
    \centering
    \includegraphics[width=0.9\linewidth]{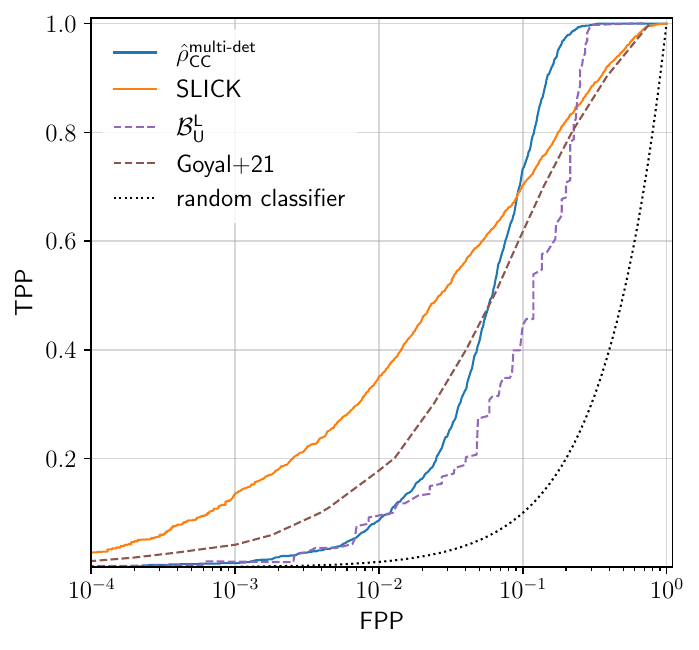}
    \caption{\label{fig:roc-comp-all}%
    ROCs from Fig.~\ref{fig:result-netw} with the addition of other methods: $\mathcal{B_\text{U}^\text{L}}$ is from the posterior overlap method~\cite{harisIdentifyingStronglyLensed2018}, while Goyal+21 corresponds to the ML method in Ref.~\cite{goyalRapidIdentificationStrongly2021}. The $\mathcal{B_\text{U}^\text{L}}$ ROC does not include the sky-localization information. Note that the dataset is different for the added dashed curves.}
\end{figure}

{\bf \em Comparing the ROC with all the existing methods---}
Fig.~\ref{fig:roc-comp-all} shows the ROC curves for a network of HLV detectors with injections having a network optimal MF SNR threshold of 8, along with individual optimal MF SNR thresholds of 4. We previously compared our result with that of Ref.~\cite{magareSLICKStrongLensing2024} in Fig.~\ref{fig:result-netw}. Here, we additionally over-plot the ROC curves for other existing methods using the same network. Particularly, we take the $(m_1^z, m_2^z)$ curve from Fig.~6(left) of Ref.~\cite{harisIdentifyingStronglyLensed2018}, which does not include sky localization information, whereas the HLV ML curve from Fig.~4 of Ref.~\cite{goyalRapidIdentificationStrongly2021} does include sky localization. Note that the datasets differ for these additional curves (dashed). Since ROCs depend on the dataset's population distribution, we do not include this plot in the main text and reserve it for reference only. Note that, in this work, we use a naive network statistic to combine the CC outputs from individual detectors. An optimal network statistic may further improve the performance.

\onecolumngrid
\clearpage

\setcounter{page}{1}
\setcounter{equation}{0}
\setcounter{figure}{0}
\setcounter{table}{0}
\setcounter{section}{0}
\setcounter{subsection}{0}
\renewcommand{\theequation}{S.\arabic{equation}}
\renewcommand{\thefigure}{S\arabic{figure}}
\renewcommand{\thetable}{S\arabic{table}}
\renewcommand{\thesection}{\Roman{section}}
\renewcommand{\thesubsection}{\Alph{subsection}}

\newcommand{\ssection}[1]{
    \addtocounter{section}{1}
    \section{\thesection.~~~#1}
    \addtocounter{section}{-1}
    \refstepcounter{section}
}
\newcommand{\ssubsection}[1]{
    \addtocounter{subsection}{1}
    \subsection{\thesubsection.~~~#1}
    \addtocounter{subsection}{-1}
    \refstepcounter{subsection}
}
\newcommand{\fakeaffil}[2]{$^{#1}$\textit{#2}\\}

\thispagestyle{empty}
\begin{center}
    \textbf{\large
        {Supplemental Material:} Optimal cross-correlation technique to search for
        \titletoken{\\}
        strongly lensed gravitational waves}\\
    \par\smallskip
    Anirban Kopty,
    Sanjit Mitra,
    and Anupreeta More
    \par
    {\small
        \fakeaffil{}{Inter-University Centre for Astronomy and Astrophysics, Pune}
    }
\end{center}
\par\smallskip

\section{Derivation of the SNR}

We start with the estimator $\hat s$ with filter $Q$, defined in the main text,
\begin{equation}
    \hat s(t) = \infint df \, \tilde s_1^*(t;f) \tilde s_2(t;f) \tilde Q^*(t;f),
\end{equation}
where we have two data strains $\{s_1, s_2\}$ consisting of GW events $\{h_1, h_2\}$ and noises $\{n_1, n_2\}$.
The signal-to-noise-ratio (SNR) would be defined by the ratio of mean of this estimator to variance of it.
\begin{itemize}
    \item \textbf{Mean:} The mean of the estimator, $\hat s(t)$ is:
    \begin{align}
        \mu_s(t) = \langle \hat s(t) \rangle
        = \infint df \, \langle \tilde s_1^*(t;f) \tilde s_2(t;f) \rangle \tilde Q^*(t;f). \label{eq:mean}
    \end{align}
    Assuming no correlation between the detectors' noise%
    ~\footnote{
        Even when dealing with two data strains in the same detector, because of a very short autocorrelation time (order of milliseconds), we can treat the two noises as uncorrelated.
    }
    and noise with signal i.e., \\ $\langle \tilde n_1^*(t;f) \tilde n_2(t;f) \rangle = \langle \tilde h_1^*(t;f) \tilde n_2(t;f) \rangle = \langle \tilde n_1^*(t;f) \tilde h_2(t;f) \rangle = 0$.
    Thus, the only surviving term is the correlation between the signals in the two detectors, i.e.,
    \begin{align}
        \mu_s(t) &= \infint df \, \langle \tilde h_1^*(t;f) \tilde h_2(t;f) \rangle \tilde Q^*(t;f), \\
        &= \infint df \, \tilde h_1^*(t;f) \tilde h_2(t;f) \tilde Q^*(t;f),
    \end{align}
    as we are dealing with deterministic signals here.

    \item \textbf{Variance:} The variance of the estimator, $\hat s(t)$ is:
    \begin{align}
        \sigma_{s}^2(t) &= \langle \hat s^2(t) \rangle - \langle \hat s(t) \rangle^2 \label{eq:variance1}.
    \end{align}
    Looking at the second term,
    \begin{align}
        \langle \hat s(t) \rangle^2 &= \langle \hat s(t) \rangle \langle \hat s^*(t) \rangle, \label{eq:variance-2nd-term} \\
        &= \infint df \infint df' \,
        \tilde h^*_1(t;f) \tilde h_2(t;f) \tilde h_1(t;f') \tilde h^*_2(t;f')
        \tilde Q^*(t;f) \tilde Q(t;f'). \label{eq:variance-2nd-term2}
    \end{align}
    And the first term,
    \begin{align}
        \langle \hat s^2(t) \rangle
        &= \langle \hat s(t) \hat s^*(t) \rangle, \label{eq:variance-1st-term} \\
        &= \left\langle \infint df \, \tilde s_1^*(t;f) \tilde s_2(t;f) \tilde Q^*(t;f) \infint df' \, \tilde s_1(t;f') \tilde s_2^*(t;f') \tilde Q(t;f') \right\rangle, \\
        &= \infint df \, \infint df' \, \tilde Q^*(t;f) \tilde Q(t;f') \left\langle \tilde s_1^*(t;f) \tilde s_2(t;f) \tilde s_1(t;f') \tilde s_2^*(t;f') \right\rangle. \label{eq:variance-1st-term2}
    \end{align}
    In Eqs. \eqref{eq:variance-2nd-term} and \eqref{eq:variance-1st-term}, we have used the property that $\hat s(t)$ is real.
    Remembering that $\tilde s_i(t;f) = \tilde h_i(t;f) + \tilde n_i(t;f)$, $i \in \{1,2\}$,
    \begin{align}
        &\left\langle
            \tilde s_1^*(t;f) \tilde s_2(t;f) \tilde s_1(t;f') \tilde s_2^*(t;f')
        \right\rangle \nonumber \\
        &= \left\langle
            (\tilde h_1^*(t;f) + \tilde n_1^*(t;f)) \, (\tilde h_2(t;f) + \tilde n_2(t;f))
            \, (\tilde h_1(t;f') + \tilde n_1(t;f')) \, (\tilde h^*_2(t;f) + \tilde n^*_2(t;f))
        \right\rangle,
    \end{align}
    where, noting that, $\left\langle \tilde n_i(t;f) \right\rangle$
    = $\left\langle \tilde n^*_1(t;f) \tilde n_2(t;f') \right\rangle$
    = $\left\langle \tilde n^*_1(t;f) \tilde n_2(t;f') \tilde n^*_2(t;f) \right\rangle$
    = 0, the only remaining terms are,
    \begin{align}
        &= \tilde h^*_1(t;f) \tilde h_2(t;f) \tilde h_1(t;f') \tilde h^*_2(t;f') \nonumber \\
        &+ \tilde h^*_1(t;f) \tilde h_1(t;f') \left\langle \tilde n_2(t;f) \tilde n^*_2(t;f') \right\rangle
        + \tilde h_2(t;f) \tilde h^*_2(t;f') \left\langle \tilde n^*_1(t;f) \tilde n_1(t;f') \right\rangle \nonumber \\
        &+ \left\langle \tilde n^*_1(t;f) \tilde n_2(t;f) \tilde n_1(t;f') \tilde n^*_2(t;f') \right\rangle, \\[3pt]
        &= \tilde h^*_1(t;f) \tilde h_2(t;f) \tilde h_1(t;f') \tilde h^*_2(t;f') \nonumber \\
        &+ \tilde h^*_1(t;f) \tilde h_1(t;f') \left\langle \tilde n_2(t;f) \tilde n^*_2(t;f') \right\rangle
        + \tilde h_2(t;f) \tilde h^*_2(t;f') \left\langle \tilde n^*_1(t;f) \tilde n_1(t;f') \right\rangle \nonumber \\
        &+ \left\langle \tilde n^*_1(t;f) \tilde n_1(t;f') \right\rangle
        \left\langle \tilde n_2(t;f') \tilde n^*_2(t;f) \right\rangle.
    \end{align}
    The noise has the following property,
    \begin{align}
        \langle \tilde n_i(t;f) \tilde n_i(t;f') \rangle
        &= \frac 12 \infint df'' S_{n,i}(t;|f''|) \delta_{\Delta t}(f''-f) \delta_{\Delta t}(f''-f'),\\
        &= \frac 12 S_{n,i}(t;|f'|) \delta_{\Delta t}(f'-f),
    \end{align}
    where we have used large chunk time limit and converted \textit{one} finite time delta function to Dirac delta function to evaluate the integral (while treating the other one as normal function). Using the above property,
    \begin{align}
        &\left\langle
            \tilde s_1^*(t;f) \tilde s_2(t;f) \tilde s_1(t;f') \tilde s_2^*(t;f')
        \right\rangle \nonumber \\
        &= \tilde h^*_1(t;f) \tilde h_2(t;f) \tilde h_1(t;f') \tilde h^*_2(t;f') \nonumber \\
        &+ \frac 12 \tilde h^*_1(t;f) \tilde h_1(t;f') S_{n,2}(t;|f'|) \delta_{\Delta t}(f'-f)
        + \frac 12 \tilde h_2(t;f) \tilde h^*_2(t;f') S_{n,1}(t;|f'|) \delta_ {\Delta t}(f'-f) \nonumber \\
        &+ \frac 14 S_{n,1}(t;|f'|) S_{n,2}(t,|f'|) \delta_{\Delta t}(f'-f) \delta_{\Delta t}(f'-f).
    \end{align}
    Putting the evaluated expression in Eq. \eqref{eq:variance-1st-term2} we get,
    \begin{align}
        \sigma^2(t) &= \langle \hat s^2(t) \rangle - \langle \hat s(t) \rangle^2, \tag{\ref{eq:variance1}} \\[5pt]
        &= \infint df \, \infint df' \, \tilde Q^*(t;f) \tilde Q(t;f')\left\langle \tilde s_1^*(t;f) \tilde s_2(t;f) \tilde s_1(t;f') \tilde s_2^*(t;f') \right\rangle \tag{\ref{eq:variance-1st-term2}} \\
        &- \infint df \infint df' \, \tilde Q^*(t;f) \tilde Q(t;f') ~
        \tilde h^*_1(t;f) \tilde h_2(t;f) \tilde h_1(t;f') \tilde h^*_2(t;f'), \tag{\ref{eq:variance-2nd-term2}} \\[5pt]
        &= \infint df \, \infint df' \, \tilde Q^*(t;f) \tilde Q(t;f')
        \bigg[
            \frac 12 \tilde h^*_1(t;f) \tilde h_1(t;f') S_{n,2}(t;|f'|) \delta_{\Delta t}(f'-f) \nonumber \\
            &+ \frac 12 \tilde h_2(t;f) \tilde h^*_2(t;f') S_{n,1}(t;|f'|) \delta_ {\Delta t}(f'-f) \nonumber \\
            &+ \frac 14 S_{n,1}(t;|f'|) S_{n,2}(t,|f'|) \delta_{\Delta t}(f'-f) \delta_{\Delta t}(f'-f)
        \bigg], \\[5pt]
        &= \infint df \, |\tilde Q(t;f)|^2
        \bigg[
            \frac 12 |\tilde h_1(t;f)|^2 S_{n,2}(t;|f|)
            + \frac 12 |\tilde h_2(t;f)|^2 S_{n,1}(t;|f|) \nonumber \\
            &+ \frac {\Delta t}4 S_{n,1}(t;|f|) S_{n,2}(t,|f|)
        \bigg], \label{eq:variance}
    \end{align}
    where again, we have transformed one finite time delta function to a Dirac delta function to evaluate the $f'$ integral, leaving the rest.

\end{itemize}
Combining the above results of mean (Eq. \ref{eq:mean}) and variance (Eq. \ref{eq:variance}) of the estimator $\hat s$, we now have the SNR,
\begin{align}
    \rho_s(t) := \frac{\mu_s(t)}{\sigma_s(t)}
    = \frac{ \infint df \, \tilde h_1^*(t;f) \tilde h_2(t;f) \tilde Q^*(t;f) }
    { \splitfrac {\bigg[ \infint df \, |\tilde Q(t;f)|^2 \times}
        {\bigg(
            \frac 12 |\tilde h_1(t;f)|^2 S_{n,2}(t;|f|)
            + \frac 12 |\tilde h_2(t;f)|^2 S_{n,1}(t;|f|)
            + \frac {\Delta t}4 S_{n,1}(t;|f|) S_{n,2}(t,|f|)
        \bigg) \bigg]^{1/2}}
    }.
\end{align}
We introduce an inner product at time t,
\begin{align}
    (\tilde A, \tilde B)_t &:= \infint df \,
    \frac{ \tilde A^*(t;f) \tilde B(t;f) }{ \xi(t;f) }, \label{eq:inner-product} \\
    \xi(t;f) &= \frac 12 |\tilde h_1(t;f)|^2 S_{n,2}(t;|f|)
        + \frac 12 |\tilde h_2(t;f)|^2 S_{n,1}(t;|f|)
        + \frac {\Delta t}4 S_{n,1}(t;|f|) S_{n,2}(t,|f|).
\end{align}
This will also define the norm of a complex function $A(t;f)$ at time t to be
\begin{equation}
    ||A||_t = \sqrt{(\tilde A, \tilde A)_t}.
\end{equation}
In terms of the newly-defined inner product, the SNR form becomes elegant:
\begin{align}
    \rho_s(t) &= \frac{(\tilde u, \tilde h_1^* \tilde h_2)_t}{(\tilde u,\tilde u)_t^{1/2}};
    \quad \tilde u(t;f) := \tilde Q(t;f) \, \xi(t;f), \\
    &= (\hat u, \tilde h_1^* \tilde h_2)_t,
\end{align}
Here, $\hat u := \tilde u/||u||_t$ is a unit vector in this space, such that $(\hat u, \hat u)=1$. Thus, we have computed the SNR $\rho_s$ for the estimator $\hat s$. In the main text, the subscript `$s$' is dropped.

\section{\texorpdfstring{Unknown time of coalescence for $s_2$}{}}

We had defined the CC statistic in the main text as,
\begin{equation}
    \rho_\text{CC} =
        \frac{(|\tilde h_1|^2, \tilde s_1^* \tilde s_2)_t}
        {(|\tilde h_1|^2, |\tilde h_1|^2)_t^{1/2}}
        e^{-im_\phi}.
    \label{eq:mf-SNR-sl}
\end{equation}
If the time of coalescence is unknown for the second event, $s_2$, one can apply a cyclical time shift in Eq. \eqref{eq:mf-SNR-sl} with $\tau$ time delay to $s_1$ (or $s_2$), i.e., $s_1(t) \rightarrow s_1(t+\tau)$ to get,
\begin{IEEEeqnarray}{lCl}
    \rho_\text{CC}(\tau)
        &=&
        \frac{(|\tilde h_1|^2, e^{2\pi i f \tau} \tilde s_1^* \tilde s_2)_t}
            {(|\tilde h_1|^2, |\tilde h_1|^2)_t^{1/2}}
            e^{-im_\phi}, \\
        &=&
        \frac{[|\tilde h_1|^2, \tilde s_1^* \tilde s_2]_\tau}
            {(|\tilde h_1|^2, |\tilde h_1|^2)_t^{1/2}}
            e^{-im_\phi},
        \label{eq:mf-SNR-tau} \\
    \rho_\text{CC}^\text{max}
        &\underset{\max \{\phi,\tau\}}{=}&
        \frac{\max \left( \text{abs} \left( [|\tilde h_1|^2, \tilde s_1^* \tilde s_2]_\tau \right) \right)}
            {(|\tilde h_1|^2, |\tilde h_1|^2)_t^{1/2}}
        \label{eq:mf-SNR-tau-maximized}
\end{IEEEeqnarray}
where we denote $[\,,]_\tau$ as the time-translated inner product operator. This operation is simply an inverse Fourier Transform, and can be efficiently computed using the fast Fourier transform (FFT) algorithm. From Eq. \eqref{eq:mf-SNR-tau} to \eqref{eq:mf-SNR-tau-maximized}, we have maximized over phase ($\phi$) by taking the absolute value and over time ($\tau$) by taking the maximum. The output can be normalized by dividing by the optimal SNR, $\rho_\text{opt}$.

\section{Trigger time being different from actual time of coalescence}

\begin{figure}
    \begin{minipage}{.49\textwidth}
        \includegraphics[width=.91\linewidth]{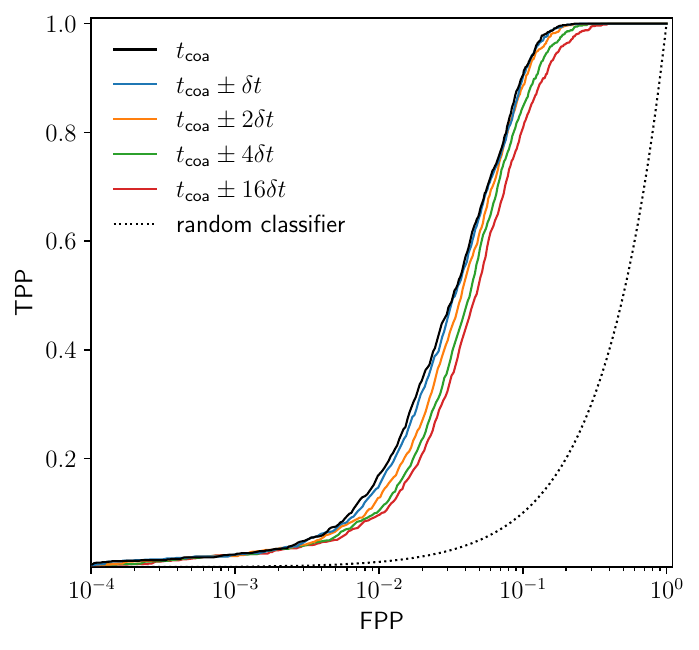}
        \caption{\label{fig:timeslide-notimeslide}%
        ROCs when the CC statistic is calculated incorporating a time lag versus when not. We find that ROC is best when we choose not to incorporate a time lag into the correlation statistic.}
    \end{minipage}%
    \hfill
    \begin{minipage}{.49\textwidth}
        \includegraphics[width=.91\linewidth]{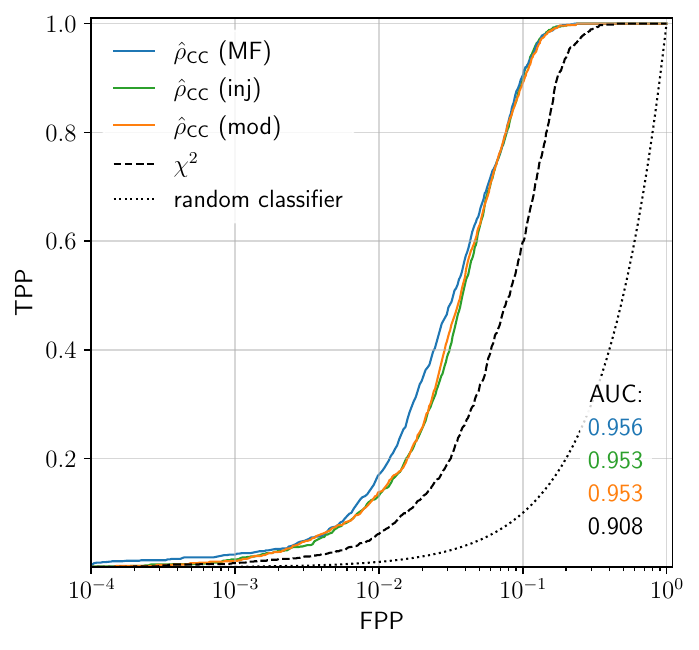}
        \caption{\label{fig:ROC-trg-inj}%
        ROC curves with AUC values for our CC statistics, when MF search is performed ($\hat \rho_\text{CC}$ (MF)) versus when not ($\hat \rho_\text{CC}$ (inj)). The differences disappear when optimal MF SNR and correct $t_\text{coa}$ is used ($\hat \rho_\text{CC}$ (mod)).}
    \end{minipage}
\end{figure}

Our CC statistic relies on the input parameters $t_\text{coa}, \tau_\text{chirp}, \tau_\text{QNM}, \mu_\text{rel}$ and $|\tilde h_1(f)|^2$. In practice, this can be taken from the MF search trigger parameters. One can understand that, out of all the above parameters, the most important parameter is the time of coalescence $t_\text{coa}$, since if it is inferred incorrectly, then we would be correlating the wrong parts of the data, where the signals will not be aligned, resulting in an unwanted value of correlation. The MF trigger time is expected not to differ significantly from the actual time of coalescence, but by only a few time samples (we have observed a maximum deviation of 4 to 6 samples).

To allow for correlating with a different $t_\text{coa}$, we can use a time lag in the CC statistic, as defined in the Appendix before. Particularly, we use Eq.~\eqref{eq:mf-SNR-tau-maximized}, maximizing over phase but not on time. After normalization, the statistic becomes,
\begin{equation}
    \hat \rho_\text{CC}(\tau)
        \underset{\max \{\phi\}}{=}
        \frac{\text{abs} \left( [|\tilde h_1|^2, \tilde s_1^* \tilde s_2]_\tau \right)}
            {(|\tilde h_1|^2, |\tilde h_1|^2)_t}
\end{equation}
We do not want to take the maximum over the full time duration, since a noise trigger can get picked up. Rather, we use the knowledge of time of coalescence and maximize over time, $\tau$ in the range $[t_\text{coa} - b\delta t, t_\text{coa} + b\delta t]$, where $b$ denotes a chosen number of time bins/samples around the known trigger time, while $\delta t$ stands for time resolution.

We perform the analysis using the same dataset as described in the main text. First, the data is searched for triggers using a matched-filter search with the same geometric non-spinning template bank as before. Then, using the trigger parameters, we compute the CC statistic with varied time lags.
To clarify, for the event with the higher MF SNR, we first cyclically shift the data's trigger time to its first bin, then we perform the CC statistic as a function of the time delay ($\tau$), and look near the second data's trigger time of coalescence.
Fig.~\ref{fig:timeslide-notimeslide} shows the ROCs when we perform our CC statistic without incorporating a time lag versus when we allow for it. We observe that optimal performance is achieved when no time lag is allowed, and the performance degrades as $b$ increases.
This is happening because, as $b$ increases, the number of noise samples (correlations with incorrect time lags) that produce spurious peaks also increases, which overwhelms any benefits derived from the correct identification of the time of coalescence.
As a result, we did not incorporate the time lag into our main analysis.

\section{ROC difference between trigger parameters and injection parameters}

We demonstrate here how the ROC curves differ due to the incorrect inference of the input parameters ($t_\text{coa}$, $\tau_\text{chirp}$, $\tau_\text{QNM}$, $\mu_\text{rel}$, $|\tilde h_1 (f)|^2$) in the matched filter search compared to the injection parameters. To reiterate on the method, we first align the two data strains using the inferred $t_\text{coa}$(s) of the two signals. We then perform \textit{time slicing} on both the data strains based on the signal width determined by $\tau_\text{chirp}$ and $\tau_\text{QNM}$. We prepare the $|\tilde h_1 (f)|^2$ by taking the trigger template and scaling its amplitude to match the MF SNR ($\rho_\text{MF}$). Finally, the CC statistic is computed using Eq.~(16) of the main text, where $\mu_\text{rel}$ is determined by $|\mu_\text{rel}| = (\rho_\text{MF}^{(2)}/\rho_\text{MF}^{(1)})^2$. For ROC with injection parameters, we use the known injection input parameters and replace the MF SNR ($\rho_\text{MF}$) with the optimal MF SNR ($\rho_\text{MF}^\text{opt}$) of the injection.

In Fig.~\ref{fig:ROC-trg-inj}, we show our result ROC (Fig.~3 in the main text) in blue, where MF denotes that MF trigger parameters are used. In green, we show the ROC for the scenario with injection parameters. $\chi^2$ result is kept in black dashed line for reference. We notice some differences in the blue and green ROC curves. Upon investigation, we find that the difference is due to two effects: incorrect inference of $t_\text{coa}$ and $\rho_\text{MF}$ being different from $\rho_\text{MF}^\text{opt}$. If we use the injection $t_\text{coa}$(s) and $\rho_\text{MF}^\text{opt}$ instead of $\rho_\text{MF}$, but keep the rest of the input parameters from trigger, then we obtain the orange curve. We find that the orange curve is similar to the green curve (AUCs are the same to 3 significant digits), indicating that other parameters do not significantly impact the results.

\end{document}